\documentclass[aps,twocolumn,showpacs,floats,prl]{revtex4}
\usepackage{graphicx}
\usepackage{dcolumn}
\usepackage{bm}
\usepackage{color}
\usepackage{amsmath}
\usepackage{subfigure}

\begin{document}

\author{N. Matveeva}
\author{S. Giorgini}
\affiliation{Dipartimento di Fisica, Universit\`a di Trento and CNR-INO BEC Center, I-38050 Povo, Trento, Italy}

\title{Fixed-node diffusion Monte Carlo study of the BCS-BEC crossover \\ in a bilayer system 
of fermionic dipoles} 

\begin{abstract} 
We investigate the BCS-BEC crossover in a bilayer system of fermionic dipoles at zero temperature using the fixed-node diffusion Monte Carlo technique. The dipoles are confined on two parallel planes separated by a distance $\lambda$ and are aligned perpendicular to the planes by an external field. The interlayer pairing, which is responsible for the superfluid behavior of the system, crosses from a weak to a strong-coupling regime by reducing the separation distance $\lambda$. For a fixed in-plane density, equal in the two layers, we calculate the ground-state energy, the chemical potential, the pairing gap and the quasiparticle dispersion as a function of the interlayer separation. At large $\lambda$ one recovers the ground-state energy of a single layer of fermions and at small $\lambda$ the one of a single layer of composite bosons with twice the particle mass and the dipole moment. The superfluid gap varies instead from the exponentially small BCS result to half of the large two-body binding energy in the BEC regime of strong interlayer pairing. Results are compared with the predictions of the simplest mean-field theory valid in the low-density limit and deviations are observed both in the BCS regime, where in-plane repulsions are important, and in the BEC regime where the mean-field approach fails to describe the physics of composite dipolar bosons.   
\end{abstract}

\maketitle 

\section{I. Introduction}

Recent progresses in experiments with polar molecules and magnetic atoms opens interesting prospects to study many-body effects in quantum degenerate gases where the dominant interactions are provided by anisotropic, long-range dipolar forces \cite{Baranov08, Baranov12}. In the case of fermionic particles, the quest for ultracold polar molecules in their rovibrational ground state is actively pursued using mixtures of $^{40}$K$^{87}$Rb \cite{Ni08, Ni10}, $^{23}$Na$^{40}$K \cite{Zwierlein}, $^{23}$Na$^6$Li \cite{Ketterle} and $^{133}$Cs$^6$Li \cite{Tiemann}. The electric dipole moment of these heteronuclear molecules ranges from 0.56 $D$ for $^{40}$K$^{87}$Rb to 5.5 D in the case of $^{133}$Cs$^6$Li. Another possibility to realize a dipolar Fermi gas is to bring highly magnetic atoms, with magnetic moments on the order of ten Bohr magnetons, to the regime of quantum degeneracy, as it has been successfully achieved with Dy~\cite{Lev} and Er~\cite{Ferlaino} atoms. 

Two dimensional (2D) geometries, in the form of a single layer or of a multilayer structure, prove to be very useful for producing ultracold gas systems with strong dipolar interactions either because they help suppress unwanted chemical reactions~\cite{Miranda11} or, more generally, they can prevent the many-body collapse driven by the head-to-tail dipolar attractive force.

We consider a bilayer system of dipolar fermions where the particles occupy two parallel planes separated by a distance $\lambda$ and the dipole moments are aligned perpendicular to the planes by a sufficiently strong external field. Tunneling between the planes is assumed to be negligible and the motion of the particles in each plane is assumed to be strictly 2D. Pairing arises from the attractive component of the interlayer dipolar interaction: two particles belonging to different layers always form a bound state for any separation distance $\lambda$~\cite{Simon76, Armstong10, Klawunn10}, while many-body systems with equal in-plane densities are expected to exhibit superfluid behavior at sufficiently low temperatures~\cite{Pikowski10, Baranov11, Zinner12}.  A crossover from a Bardeen-Cooper-Schrieffer (BCS) to a Bose-Einstein condensate (BEC) type of superfluid state is also expected as a function of the interlayer distance, depending on whether the two-body binding energy is smaller or larger than the in-plane Fermi energy. A qualitative description of this crossover is provided by the BCS theory applied to 2D Fermi gases~\cite{Miyake83, Randeria90}.

The bilayer system of fermionic dipoles considered in the present article shows a novel type of BCS-BEC crossover similar to the one studied in two component Fermi gases where contact interactions are tuned by a magnetic field in the vicinity of a Fano-Feschbach resonance (for a review see Refs.~\cite{Giorgini, Bloch}). A new ingredient here is the long-range nature of the dipolar interaction and the in-plane repulsion felt by the particles. These latter features establish also strong analogies with the electron-hole bilayer in semiconductor heterostructures and graphene, which attracted a lot of interest in the last years~\cite{Senatore, Needs, Perali}. 

Previous theoretical studies were performed in the mean-field approximation~\cite{Pikowski10, Baranov11, Zinner12}. Here we report on calculations of the equation of state and of the superfluid gap at zero temperature using the fixed-node diffusion Monte Carlo (FN-DMC) technique. The in-plane density corresponds to the weakly-interacting regime of a single-layer Fermi liquid~\cite{Shlyapnikov12, Matveeva12}. For balanced populations in the two layers we calculate the ground-state energy of the system and, from the dependence of this energy on a slight population unbalance, we determine the chemical potential and the pairing gap. By decreasing the interlayer separation $\lambda$ the ground-state energy varies from the value corresponding to a single fermionic layer~\cite{Matveeva12} to the one of a single layer of composite bosonic dipoles with twice the mass and twice the dipole moment. In the same crossover, the pairing gap increases from the exponentially small BCS result to half of the large two-body binding energy in the BEC regime of small separation. We compare our results with the simplest mean-field approach valid in the low-density limit and we find important deviations once the contribution from the two-body physics is subtracted from the energy per particle and the pairing gap. The role played by in-plane repulsions is also found to be relevant in the discussion of the schematic phase diagram of the system in the interlayer/intralayer interaction plane, where the BCS and BEC regimes of the superfluid compete with the Wigner crystal phase reached at large densities.  

The structure of the paper is as follows: in Sec. II we describe our model Hamiltonian, provide some basic information about the FN-DMC technique and discuss the choice of the trial wave function used in the calculations. Sec. III contains a review of the results of the mean-field approach, which qualitatively describes the crossover in terms of the two-body binding energy. In Sec. IV we present the FN-DMC results for the ground-state energy and we discuss the phase diagram of the system as a function of interlayer separation and in-plane interaction strength. In Sec. V the technique to calculate the chemical potential, the pairing gap and the quasiparticle spectrum is explained and the results are discussed and compared with mean-field predictions. Finally, we draw our conclusions.

\section{II. Model and FN-DMC method}
\begin{figure}
\begin{center}
\includegraphics[width=6.0cm]{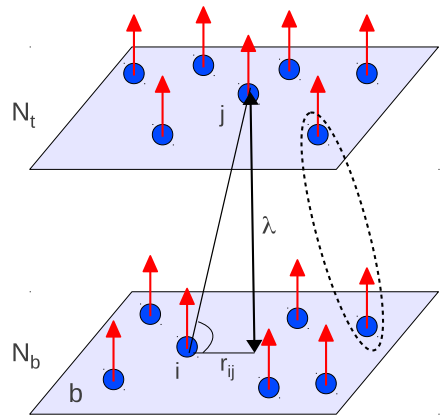}
\caption{(color online). Schematic view of the bilayer system of dipolar fermions. 
\label{fig1}}
\end{center}
\end{figure}

We consider a bilayer system of identical fermionic dipoles where bottom and top layers contain, respectively, $N_b=N/2$ and $N_t=N/2$ particles, $N$ being the total number of fermions. The layers are strictly 2D planes separated by a distance $\lambda$. We assume that all dipoles are aligned perpendicular to the plane of motion by an external field and, also, that tunneling between layers can be neglected.  The Hamiltonian of such a system is written as
\begin{eqnarray}
H&=&-\frac{\hbar^2}{2m}(\sum_{i=1}^{N_b}\nabla_i^2+\sum_{j=1}^{N_t}\nabla_j^2) \\
&+&\sum_{i<i^{\prime}}^{N_b}V_b(r_{ii^{\prime}}) +\sum_{j<j^{\prime}}^{N_t}V_t(r_{jj^{\prime}})+ \sum_{i,j}^{N_b, N_t} V_{int}(r_{ij}) \;.
\nonumber
\label{Hamiltonian} 
\end{eqnarray}
Here $m$ denotes the mass of the particles, $d$ is the dipole moment, $r_{ii^\prime}$ and $r_{jj^\prime}$ denote, respectively, the in-plane interparticle distance in the bottom layer between the $i$-th and the $i^\prime$-th particle and in the top layer between the $j$-th and the $j^\prime$-th particle. The in-plane interaction potential in the bottom (top) layer, $V_{b(t)}$, is purely repulsive and is given by $V_{b(t)}(r) = d^2/r^3$. The interlayer potential $V_{int}$, instead, is given by the formula
\begin{equation}
V_{int}(r_{ij}) = \frac{d^2(r_{ij}^2 - 2 \lambda^2)}{(r_{ij}^2+\lambda^2)^{5/2}}\;,
\label{V_inter} 
\end{equation} 
where $r_{ij}=|{\bf r_i} -{\bf r_j}|$ is the in-plane distance between the $i$-th particle in the bottom layer and the projection onto the bottom layer of the position of the $j$-th particle in the top layer  (see Fig.~\ref{fig1}). The strength of the in-plane and the interlayer dipolar interaction is described
in terms of the dimensionless parameter $k_Fr_0$ and $k_F\lambda$, respectively. Here $k_F=\sqrt{4\pi n_{sl}}$ is the Fermi wave vector determined by the density $n_{sl}$ in each layer and $r_0=md^2/\hbar^2$ is the characteristic length of the dipolar potential. It is important to stress that the potential (\ref{V_inter}), for any value of the interlayer distance $\lambda$, sustains a two-body bound state with energy $E_b$~\cite{Simon76, Armstong10, Klawunn10}.

As anticipated in the Introduction, we use the FN-DMC method in order to calculate various ground-state properties of the system~\cite{Kolorenc11}. The method is based on the choice of a trial wave function which, for fermions, must be antisymmetric with respect to the exchange of identical particles. FN-DMC simulations provide a rigorous upper bound to the ground-state energy depending on the choice of the nodal surface of the trial wave function, {\it i.e.} the multidimensional surface in configuration space where the many-body wave function vanishes. In principle, if the nodal surface of the trial wave function is exact, the FN-DMC estimate of the ground-state energy is also exact. 

Simulations are carried out in a box of volume $\Omega = L^2$ with the single-layer density $n_{sl}=\frac{N}{2L^2}$. Periodic boundary conditions (PBC) are used in both spatial directions. In order to account correctly for the long-range character of the interaction energy we use a numerical procedure
equivalent to the Ewald's summation technique~\cite{Kolorenc11}, but in our case the sums are evaluated in real space (see Appendix A).

The trial wave function used to impose the nodal surface constraint is given by
\begin{equation}
\Psi_T({\bf R})= \prod_{i<i^{\prime}}^{N_b}f(r_{ii^\prime})  \prod_{j<j^{\prime}}^{N_t}
f(r_{jj^\prime}) \Psi_A({\bf R}) \;,
\label{Trial} 
\end{equation}
where ${\bf R}=({\bf r}_1,\dots,{\bf r}_{2N})$ is the multidimensional vector denoting the spatial coordinates of the particles. The function $f(r)$ is a two-body non-negative Jastrow term describing in-plane correlations. It is parametrized as $f(r)\propto K_0(2\sqrt{r_0/r})$ for $r<\bar{R}$ and $f(r)\propto \exp(-C/r)$ for $r>\bar{R}$, where $K_0$ is the modified Bessel function, $C$ is a constant determined through the condition $f^\prime(r=L/2) = 0$ and $\bar{R}$ is a variational parameter~\cite{Matveeva12}. The term $\Psi_A({\bf R})$ is chosen as the antisymmetrized product of pairwise orbitals $\phi({\bf r}_{ij})$:
\begin{equation}
\Psi_A({\bf R})= \det 
\begin{bmatrix}
\phi({\bf r}_{11}) & \cdots &  \phi({\bf r}_{1N_t})  \\
\vdots                     & \ddots &     \vdots \\
\phi({\bf r}_{N_b1}) & \cdots &  \phi({\bf r}_{N_bN_t})  \\
\end{bmatrix} \;.
\label{PsiA}
\end{equation}
The orbitals are taken of the general form
\begin{equation}  
\phi({\bf r})=A\; h(r)+B\;\sum_{k_{\alpha}=0}^{k_F}\exp[i{\bf k_{\alpha}}\cdot{\bf r}],
\label{phi}
\end{equation}
where ${\bf k_{\alpha}} = (2\pi/L)(n^x_{\alpha},n^y_{\alpha})$ are the wave vectors complying with PBC in the box of size $L$ and $A$ and $B$ are variational parameters. The function $h(r)$ is parametrized as
\begin{equation}
h(r) = e^{-\gamma \sqrt{r^2/\lambda^2 +1}} + e^{\gamma\left(\sqrt{r^2/\lambda^2+1} -2 \sqrt{L^2/4\lambda^2+1}\right)} \;,
\label{h}
\end{equation}
with $\gamma$ a variational parameter. At small interlayer separations $\lambda$ the functional form of $h(r)$ agrees, for small distances $r$, with the lowest two-body bound state of the potential (\ref{V_inter}). We also notice that the value of the parameters $A$, $B$ and $\gamma$ present in the many-body wave function (\ref{PsiA}) modify the nodal surface and therefore require an optimization procedure using the FN-DMC algorithm.

The pair orbital (\ref{phi}) provides the correct description of two important regimes. The first one corresponds to $B=0$, in which case $\Psi_A$ describes an antisymmetric state of composite bosonic dimers. The second regime corresponds to $A=0$: here $\Psi_A$ is equal to the product of the plane-wave Slater determinants for bottom and top layer particles
$\Psi_A({\bf R})=\det[e^{i{\bf k_{\alpha}}{\bf r_i}}]\times\det[e^{i{\bf k_{\alpha}}{\bf r_j}}]$.  In this latter case the nodal surface of the trial wave function coincides with the one of an ideal Fermi gas and, being incompatible with off-diagonal long-range order in the two-body density matrix~\cite{Astra05}, properly describes normal Fermi liquids. This choice of the trial wave function is expected to be valid in the deep BCS regime, where the effects of pairing on the ground-state energy are negligible. The parametrization (\ref{phi}) of the pair orbital allows one to interpolate continuously between these two regimes~\cite{Chang04, Astra05}.

\section{III. Mean-field results}
As it is known from BCS theory \cite{Miyake83, Randeria90}, in two dimensions the presence of a two-body bound state in vacuum is a necessary and sufficient condition for s-wave pairing with an arbitrary interaction potential. The solution of the BCS equations is particularly simple in 2D providing the following analytical results at $T=0$:
\begin{equation}
\Delta=\sqrt{2\epsilon_F|E_b|} \;,
\label{Delta}
\end{equation}
and
\begin{equation}
\mu=\epsilon_F + E_b/2 \;,
\label{mu_BCS}
\end{equation}
for the order parameter and the chemical potential respectively. These results, which only involve the energy $E_b$ of the two-body bound state and the Fermi energy $\epsilon_F=\hbar^2k_F^2/(2m)$, can be applied to the bilayer system in the low-density limit where in-plane interactions and anomalous contributions to inter-layer scattering are both negligible~\cite{Pikowski10,Baranov11}. 
More sophisticated mean-field approaches have been developed~\cite{Zinner12} that incorporate 
interaction effects beyond the dilute limit, but they rely on full numerical solutions of the BCS equations. From the thermodynamic relation $\mu=dE/dN$ one gets from Eq.~(\ref{mu_BCS}) the following result for the energy per particle in the ground state
\begin{equation}
\frac{E}{N}=E_{IFG} + E_b/2 \;,
\label{en_BCS}
\end{equation}
where $E_{IFG}=\epsilon_F/2$ is the energy per particle of a noninteracting gas. Quasiparticle excitations above the ground state are described within the BCS theory by the dispersion relation
\begin{equation}
\epsilon_k = \sqrt{\left(\frac{\hbar^2k^2}{2m}-\mu\right)^2 + \Delta^2} \;,
\label{Dispersion}
\end{equation}
and the pairing gap $\Delta_{gap}$ is defined as $\Delta_{gap}=\min_k (\epsilon_k)$. In the BCS regime, where $\mu>0$, the excitation energy $\epsilon_k$ has the minimum at $k=\sqrt{2m\mu/\hbar^2}$ and the pairing gap coincides with the order parameter: $\Delta_{gap}=\Delta$. In the BEC regime, where $\mu<0$, the dispersion relation (\ref{Dispersion}) has its minimum at $k=0$ and in this case
\begin{equation}
\Delta_{gap}=\sqrt{\mu^2+\Delta^2} \;.
\label{GAP_BEC}
\end{equation}
By substituting the chemical potential from Eq.~(\ref{mu_BCS}) into Eq.~(\ref{GAP_BEC}) one obtains $\Delta_{gap}=\epsilon_F+|E_b|/2$ for the pairing gap in this regime. The above mean-field predictions will be used in the following sections to provide a comparison with the results of FN-DMC simulations.

\section{IV. Ground-state energy}
\begin{figure}
\begin{center}
\includegraphics[width=8.5cm]{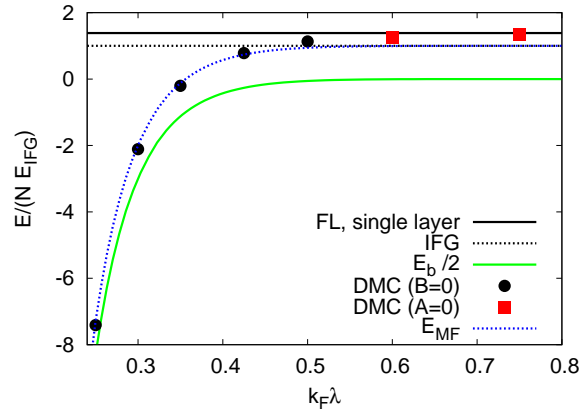}
\caption{(color online). Ground-state energy as a function of the interlayer distance $k_F\lambda$. Symbols refer to FN-DMC calculations using the trial wave function (\ref{Trial}) with $B=0$ in Eq.~(\ref{phi}) (black circles) and $A=0$ in Eq.~(\ref{phi}) (red squares). Lines are as follows: noninteracting Fermi gas (horizontal black dotted), mean-field theory (blue dotted), half of two-body binding energy (green solid) and single-layer Fermi liquid (horizontal black solid). 
\label{fig2}}
\end{center}
\end{figure}

\begin{figure}
\begin{center}
\includegraphics[width=8.5cm]{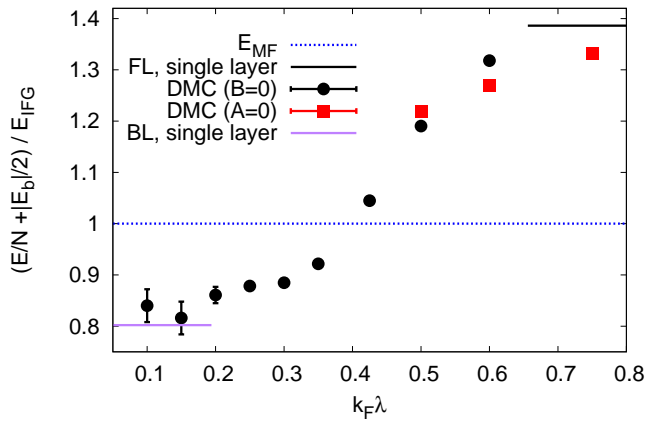}
\caption{(color online). Ground-state energy with $E_b/2$ subtracted as a function of the interlayer distance $k_F\lambda$. Symbols refer to FN-DMC calculations using the trial wave function (\ref{Trial}) with $B=0$ in Eq.~(\ref{phi}) (black circles) and $A=0$ in Eq.~(\ref{phi}) (red squares). Lines are as follows: mean-field theory (blue dotted), single-layer Fermi liquid (top black solid) and single-layer composite bosons (bottom purple solid). 
\label{fig3}}
\end{center}
\end{figure}

In this section we discuss the FN-DMC results obtained for the ground-state energy as a function of the dimensionless interlayer distance $k_F\lambda$ (see Figs.~\ref{fig2} and \ref{fig3}). The in-plane interaction strength is taken as $k_Fr_0=0.5$ corresponding, in the case of a single layer, to a weakly interacting Fermi liquid~\cite{Matveeva12}. Calculations are performed using two wave functions both parametrized by Eqs.~(\ref{Trial}), (\ref{PsiA}): the first contains the pair orbital $\phi(r)$ of Eq.~(\ref{phi}) with $B=0$ and corresponds to a BCS-type wave function of composite bosonic dimers, the second contains $\phi(r)$ with $A=0$ and is equivalent to a Jastrow-Slater wave function. Finite-size errors are analyzed following the procedure described in Appendix B and the results extrapolated to the thermodynamic limit are shown in Figs.~\ref{fig2} and \ref{fig3}. The BCS-type wave function ($B=0$) is found to give lower energies for $k_F\lambda \lesssim 0.5$ (see Fig.~\ref{fig3}).

We compare our FN-DMC data with the result of the mean-field theory from Eq.~(\ref{en_BCS})
(blue dashed line in Fig.~\ref{fig2}). One can see that if $E_b/2$ is not subtracted from $E/N$ there appear to be good agreement between mean-field and FN-DMC results, especially at small values of $k_F\lambda$ where the two-body contribution (shown in Fig.~\ref{fig2} as a green solid line) dominates over the many-body contribution. For large interlayer distances, the energy of a single-layer Fermi liquid, given by $E/N=1.3862(5) E_{IFG}$ \cite{Matveeva12}, is almost exactly recovered. 

Once the binding energy contribution is subtracted from $E/N$ (see Fig.~\ref{fig3}), deviations are visible compared to Eq.~(\ref{en_BCS}) (shown as a blue dashed line in Fig.~\ref{fig3}). 
At relatively large values of $k_F\lambda$ it is evident that the energy approaches the value of the
single-layer interacting gas and this effect is completely not accounted for by the mean-field result (\ref{en_BCS}). In the opposite regime of small $k_F\lambda$, one should compare $E/N-E_b/2$ with the energy of dipolar composite bosons having 
mass $2m$, dipolar strength $2d$ and dipolar length $\tilde{r}_0=8r_0$. At the effective density $n_{sl}\tilde{r}_0^2=1.27$, corresponding to $k_Fr_0=0.5$ for single-layer fermions, these composite bosons have an energy per particle $E/N=0.8021(3) E_{IFG}$ 
(shown in Fig.~\ref{fig3} as an bottom purple solid line) \cite{Note}. We see that by reducing $k_F\lambda$ the FN-DMC energies approach this asymptotic value, showing that energy-wise the system indeed behaves as a single layer of composite bosons interacting with a much larger dipolar strength ($n_{sl}\tilde{r}_0^2=64n_{sl}r_0^2$). We notice that in the region $0.5\lesssim k_F\lambda\lesssim 0.6$, where the nodal constraint of the BCS-type wave function becomes energetically favorable over the one of the Jastrow-Slater wave function, one expects that a more advanced nodal surface, interpolating between the two limits, may provide a lower bound for the ground-state energy. 

\begin{figure}
\begin{center}
\includegraphics[width=8.5cm]{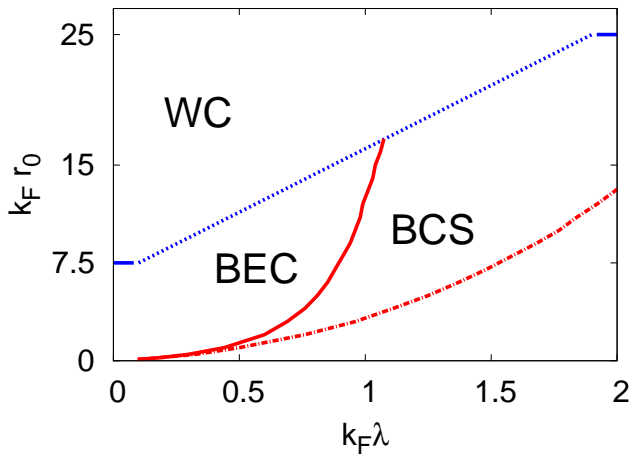}
\caption{(color online). Schematic phase diagram in the plane spanned by $k_F\lambda$ and 
$k_Fr_0$. The blue dotted line indicates in a qualitative way the separation between the Wigner crystal and the superfluid phase within the known limits of a single layer of dipolar fermions (large $k_F\lambda$) and of a single layer of dipolar composite bosons (small $k_F\lambda$). The red solid line separates the BEC from the BCS region in the superfluid phase, respectively defined as
where $\mu_{sl}<|E_b|/2$ and $\mu_{sl}>|E_b|/2$, in terms of the chemical potential $\mu_{sl}$ of a single layer of dipolar fermions. The red dashed line shows the BEC-BCS separation when 
$\mu_{sl}=\epsilon_F$, valid in the low-density limit.  
\label{fig3-1}}
\end{center}
\end{figure}

In Fig.~\ref{fig3-1} we show a schematic plot of the phase diagram of the bilayer system as a function of the parameters $k_Fr_0$ and $k_F\lambda$. For a large interaction strength $k_Fr_0$, the system is expected to enter the Wigner crystal (WC) phase at any value of the dimensionless
inter-layer distance $k_F\lambda$. In particular, for large $k_F\lambda$, the critical density
where the fluid to solid transition occurs should coincide with the one of a single layer of dipolar fermions, {\it i.e.} $k_Fr_0=25(3)$ as it has been obtained in Ref.~\cite{Matveeva12}. In Fig.~\ref{fig3-1} we arbitrarily assumed that at $k_F\lambda=2$ the transition point is already close to this critical value. This picture is supported by the results on the equation of state reported in Fig.~\ref{fig2}, where the ground-state energy of the bilayer and the single layer of dipolar fermions agree well for $k_F\lambda>1$. In the opposite regime, $k_F\lambda\ll 1$, the bilayer system of fermions behaves as a single layer of dipolar composite bosons with dipole moment $2d$ and mass $2m$, corresponding to the effective dipolar length $\tilde{r}_0=8r_0$. The superfluid to solid transition of a single layer of dipolar bosons was investigated in Ref.~\cite{Astra07} and the critical value $n_{sl}\tilde{r}_0^2=290(30)$ was obtained. This value converts into $k_Fr_0=7.5(8)$, as reported in Fig.~\ref{fig3-1} where a straight line connects the two known limits providing a qualitative picture of the phase diagram. In the same figure, the superfluid region below the blue dotted line is separated into a BEC and a BCS part, which are approximately established as where the single-layer chemical potential $\mu_{sl}<|E_b|/2$ and $\mu_{sl}>|E_b|/2$, respectively. The value of $\mu_{sl}$ is derived from the results of the ground-state energy $E_{sl}$, obtained in Ref.~\cite{Matveeva12} for a single layer of dipolar fermions, using the thermodynamic relation $\mu_{sl}=dE_{sl}/dN$. In Fig.~\ref{fig3-1} we also show the BEC-BCS separation when $\mu_{sl}=\epsilon_F$, extrapolating from the dilute limit. The large reduction of the BEC region in the case of the full determination of $\mu_{sl}$ is mainly due to the in-plane repulsion which increases the value of the chemical potential in agreement with the findings of Ref.~\cite{Zinner12}.

\section{V. Pairing gap, chemical potential and quasiparticle excitation spectrum}
\subsection{Method}
In order to calculate using the FN-DMC method the pairing gap $\Delta_{gap}$ and the chemical potential $\mu$ we need to consider a polarized system with $N_t>N_b$. In this case the antisymmetric many-body wave function (\ref{PsiA}) should be generalized to deal with a number of only $N_b$ pairwise orbitals and the remaining number $M=N_t-N_b$ of unpaired particles occupying single-particle states. We use the following form of $\Psi_A$ that has already been successfully employed in the study of polarized systems of fermions~\cite{Carlson03}
\begin{equation}
\Psi_A({\bf R})= \det 
\begin{bmatrix}
\phi({\bf r}_{11}) & \cdots &  \phi({\bf r}_{1N_t})  \\
\vdots                     & \ddots &     \vdots \\
\phi({\bf r}_{N_b1}) & \cdots &  \phi({\bf r}_{N_bN_t})  \\
\varphi_1({\bf r}_1) & \cdots & \varphi_1({\bf r}_{N_t}) \\
\vdots                     & \ddots &     \vdots \\
\varphi_M({\bf r}_1) & \cdots & \varphi_M({\bf r}_{N_t}) \\
\end{bmatrix} \;.
\label{PsiApol}
\end{equation}
The pairwise orbitals $\phi({\bf r})$ are chosen of the form (\ref{phi}), with the function $h(r)$ given by Eq.~(\ref{h}). A simple choice of the single-particle states $\varphi_{\alpha}({\bf r})$ is provided by the plane waves complying with PBC in the box of size $L$. The wave vectors ${\bf k}_\alpha$, $\alpha=1,...,M$, are chosen such that the nodal surface of the many-body wave function (\ref{PsiApol}) is the one of minimal energy. We restrict our calculation of $\Delta_{gap}$ and $\mu$ to the values of $k_F\lambda\leq0.5$, where the choice of $B=0$ for the pairwise orbitals in Eq.~(\ref{phi}) gives the lowest energy.  

\begin{figure}
\begin{center}
\includegraphics[width=9cm]{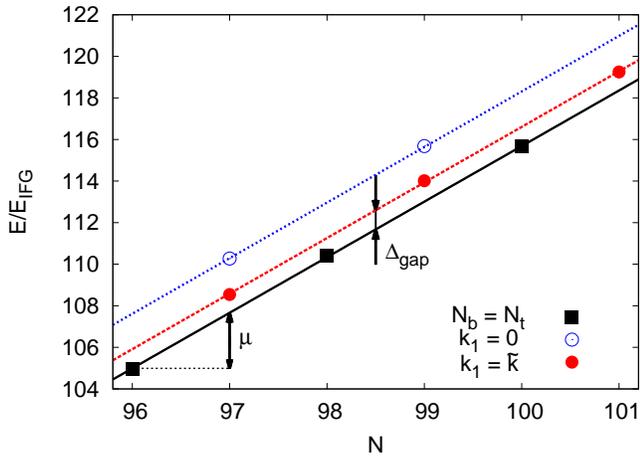}
\caption{(color online). Dependence of the total energy $E$ on the number of particles $N$ in the two layers at $k_F\lambda =0.5$. Black squares refer to the balanced case $N_b=N_t$.  Red solid and blue empty circles are, respectively, the energy for the unbalanced case $N_t=N_b+1$ where the unpaired particle has wave vector $k_1=\tilde{k}$ and $k_1=0$. Lines are linear fits through the data.  
\label{fig4}}
\end{center}
\end{figure}

We determine $\Delta_{gap}$ and $\mu$ from the following relation between the energy of the balanced system and the system with one extra particle in the top layer
\begin{equation}
E\left(\frac{N}{2}+1, \frac{N}{2}\right)=E\left(\frac{N}{2},\frac{N}{2}\right)+\mu + \Delta_{gap} \;.
\label{GAP}
\end{equation}
Here $E(N/2,N/2)$ is the ground-state energy of the system with $N/2$ particles in each layer and $E(N/2+1,N/2)$ is the ground-state energy of the system with $N/2+1$ particles in the top layer and $N/2$ particles in the bottom layer. In order to calculate the energy $E(N/2+1,N/2)$ we make use of the trial function (\ref{PsiApol}) with a single unpaired particle $M=1$. The corresponding orbital $\varphi_1({\bf r})=\cos({\bf k}_1\cdot{\bf r})$ can be easily optimized by choosing ${\bf k}_1$ among the wave vectors ${\bf k_{\alpha}} = (2\pi/L)(n^x_{\alpha},n^y_{\alpha})$ complying with PBC. Fig.~\ref{fig4} shows the results of the calculation of $\Delta_{gap}$ and $\mu$ at $k_F\lambda=0.5$. Two values of $k_1$ are reported for comparison: $k_1=0$ and $k_1=\tilde{k}=4(2\pi/L)$, the latter giving the lowest energy $E(N/2+1,N/2)$. The energies of the balanced and polarized systems depend linearly on $N$ and for both the slope is given by $\mu$. The pairing gap $\Delta_{gap}$ is obtained as the vertical distance between the lines used to fit the energies of the balanced and polarized systems. In all calculations reported in Fig.~\ref{fig4} the size $L$ of the simulation box is kept fixed and equal to $L=\sqrt{N/2n_{sl}}$,  where $N=98$ and $n_{sl}$ is the single-layer density such that $k_Fr_0=0.5$. We notice that the value of $\tilde{k}$, which minimizes $\Delta_{gap}$ in Fig.~\ref{fig4}, is the wave vector ${\bf k}_\alpha$ closer to $k_F$. 

The results of Fig.~\ref{fig4}, and other results of $\mu$ and $\Delta_{gap}$ obtained using Eq.~(\ref{GAP}) for different values of $k_F\lambda$, are shown in the next section. At the largest interlayer separations the optimal ${\bf k}_1$ remains close to the Fermi wave vector $k_F$, whereas at small separations ${\bf k}_1=0$ gives the lowest energy. In this latter regime, the method outlined above to calculate $\Delta_{gap}$ becomes less accurate and we resort to another relation defining the gap
\begin{equation}
E\left(\frac{N}{2}+\frac{M}{2},\frac{N}{2}-\frac{M}{2}\right)=E\left(\frac{N}{2},\frac{N}{2}\right) + M \Delta_{gap} \;,
\label{GAP_1}
\end{equation}
which holds in the limit $M\ll N$. In terms of the polarization $P=(N_t-N_b)/(N_t+N_b)=M/N$ the above equation becomes 
\begin{equation}
\frac{E(P)}{N}=\frac{E(P=0)}{N} + \Delta_{gap}P \;.
\label{GAP_1_1}
\end{equation}

\begin{figure}
\begin{center}
\includegraphics[width=9cm]{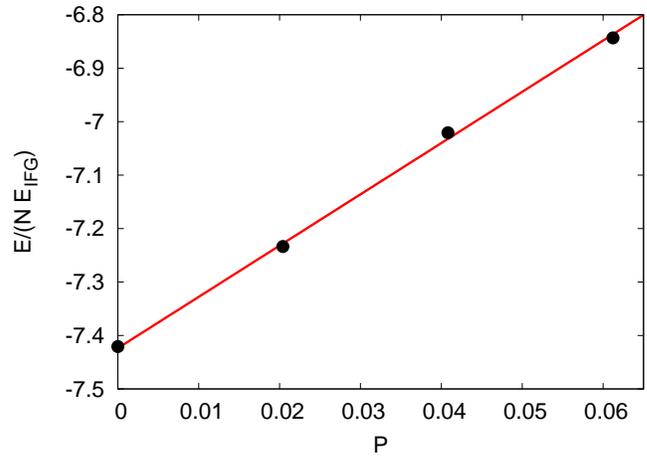}
\caption{(color online).  Dependence of the ground-state energy $E$ on the polarization $P$ at $k_F\lambda =0.25$. The total number of particles is $N=98$. 
\label{fig5}}
\end{center}
\end{figure}

In Fig.~\ref{fig5} we show the results of $E(P)$ at the separation distance $k_F\lambda=0.25$. Here we calculate the ground-state energy for $M=0,2,4,6$, with the fixed total number of particles $N=98$. As for the calculation reported in Fig.~\ref{fig4}, the size of the simulation box is fixed to $L=\sqrt{N/2n_{sl}}$. The unpaired particles occupy, starting from the first, the following set of plane-wave states: $\varphi_1({\bf r})=1$, $\varphi_2({\bf r})=\cos({\bf k_1\cdot r})$, $\varphi_3({\bf r})=\sin({\bf k_1\cdot r})$, $\varphi_4({\bf r})=\cos({\bf k_2\cdot r})$, $\varphi_5({\bf r})=\sin({\bf k_2\cdot r})$, $\varphi_6({\bf r})=\cos({\bf k_3\cdot r})$, where ${\bf k_1} = \frac{2\pi}{L}(1,0)$, ${\bf k_2} = \frac{2\pi}{L}(0,1)$ and ${\bf k_3} = \frac{2\pi}{L}(1,1)$. At the same separation distance, $k_F\lambda=0.25$, we also calculated the gap using Eq.~(\ref{GAP}) finding agreement between the two methods within statistical uncertainty. The advantage of the method based on Eq.~(\ref{GAP_1_1}) is the higher precision when the value of $\Delta_{gap}$ is large compared to the Fermi energy $E_F$. For this reason we make use of Eq.~(\ref{GAP_1_1}) at interlayer separations $k_F\lambda\le0.25$ and the corresponding results are presented in the next section.

We also checked finite-size errors by carrying out calculations with $N= 26, 58, 98$ finding all corresponding values of $\Delta_{gap}$ and $\mu$ in agreement within error bars.

\subsection{Results}
In this section we discuss the main results for the pairing gap, the chemical potential and the excitation spectrum, comparing them with mean-field predictions.

\begin{figure}
\begin{center}
\includegraphics[width=9cm]{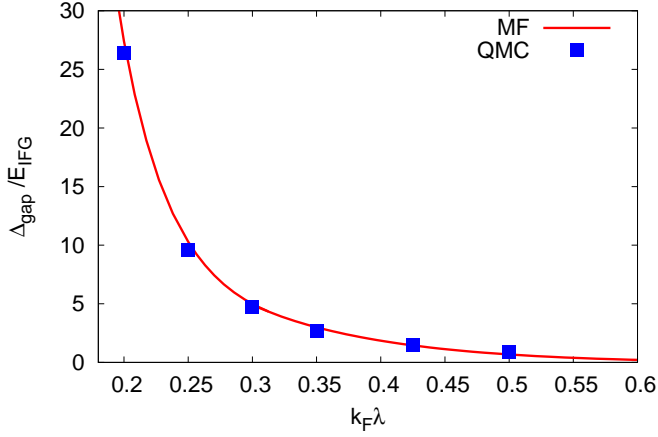}
\caption{(color online). Pairing gap as a function of $k_F\lambda$. Blue squares are the FN-DMC results while the line is the result of mean-field theory.
\label{fig6}}
\end{center}
\end{figure}

\begin{figure}
\begin{center}
\includegraphics[width=9cm]{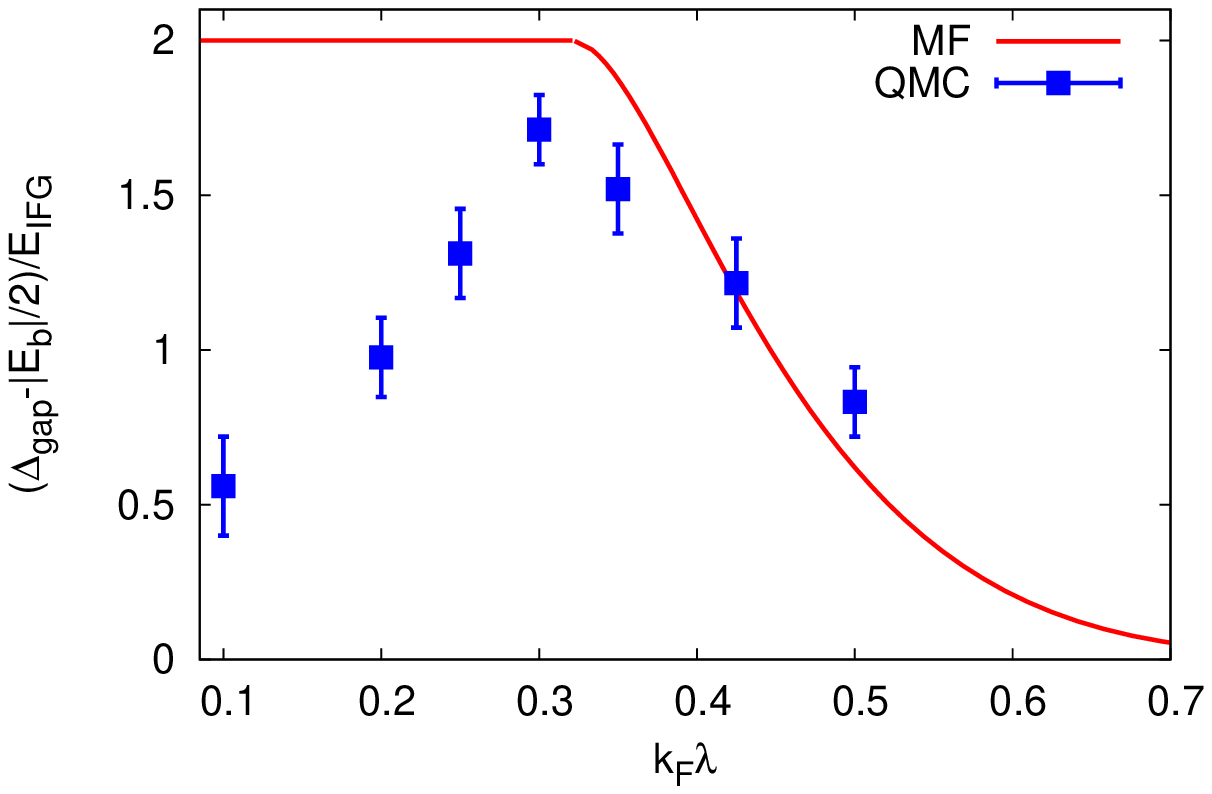}
\caption{(color online). Pairing gap as a function of $k_F\lambda$ with $|E_b|/2$ subtracted. Blue squares are the FN-DMC results while the line is the result of mean-field theory.  
\label{fig7}}
\end{center}
\end{figure}

\begin{figure}
\begin{center}
\includegraphics[width=9cm]{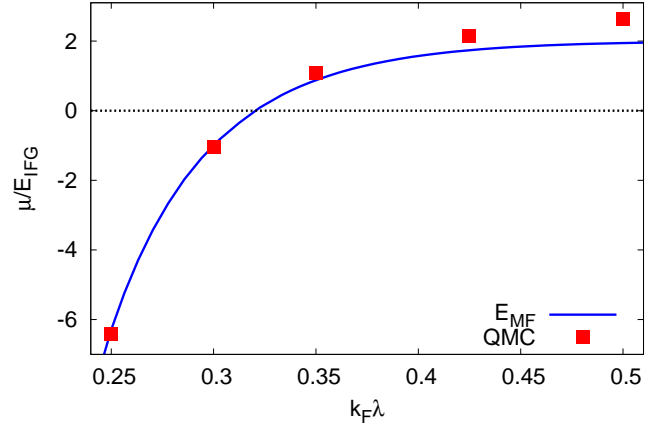}
\caption{(color online). Chemical potential as a function of $k_F\lambda$. Red squares are the FN-DMC results while the line is the result of mean-field theory.  
\label{fig8}}
\end{center}
\end{figure}

First we compare the FN-DMC results for $\Delta_{gap}$ and $\mu$ with Eqs.~(\ref{GAP_BEC}) and (\ref{mu_BCS}) respectively. In Fig.~\ref{fig6} we show the pairing gap as a function of $k_F\lambda$, without subtracting $|E_b|/2$ and we find good agreement with mean-field theory. However, once the trivial contribution from the two-body bound state is subtracted (see Fig.~\ref{fig7}), significant deviations become visible especially in the BEC regime where the mean-field theory does not account for effects related to the dimer-dimer interaction. The results for the chemical potential are shown in Fig.~\ref{fig8} for the values of $k_F\lambda$ where we employed Eq.~(\ref{GAP}) to determine $\Delta_{gap}$. We notice that at small separations the agreement with the mean-field result (\ref{mu_BCS}) is good, consistently with the findings for the energy per particle reported in Fig.~\ref{fig2}. For the largest values of $k_F\lambda$, the mean-field prediction does not account for the in-plane repulsion and lies significantly below the FN-DMC result. Both FN-DMC and mean-field results indicate that the chemical potential changes sign at $k_F\lambda \sim0.325$ (see also Ref.~\cite{Zinner12}).

\begin{figure}
\begin{center}
\includegraphics[width=9cm]{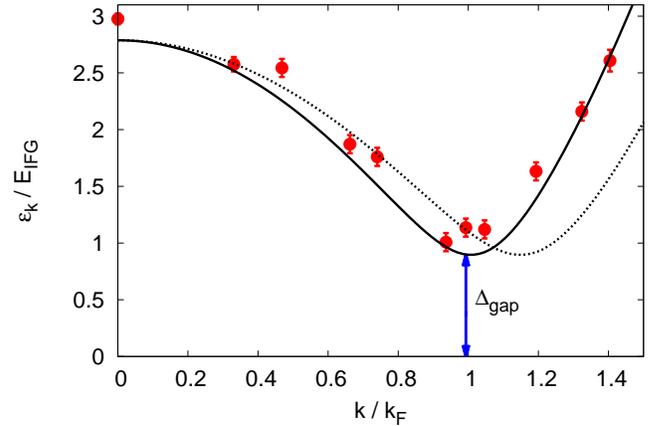}
\caption{(color online). The excitation spectrum in the BCS regime at $k_F\lambda = 0.5$. The red symbols are the FN-DMC results, the dotted line is the spectrum (\ref{Dispersion}) and the solid line is the modified dispersion (\ref{Dispersion_1}) where $m^\star$ is a fitting parameter. In both Eq.~(\ref{Dispersion}) and Eq.~(\ref{Dispersion_1}) we use the FN-DMC values for $\Delta_{gap}$ and $\mu$.
\label{fig9}}
\end{center}
\end{figure}

\begin{figure}
\begin{center}
\includegraphics[width=9cm]{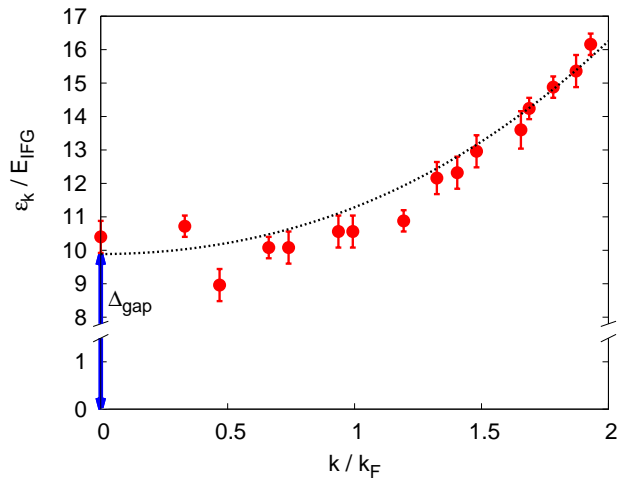}
\caption{(color online). The excitation spectrum in the BEC regime at $k_F\lambda = 0.25$. The red symbols are the FN-DMC results and the dotted line is the spectrum (\ref{Dispersion}) where we use the FN-DMC values for $\Delta_{gap}$ and $\mu$.  
\label{fig10}}
\end{center}
\end{figure}

In Figs.~\ref{fig9} and \ref{fig10} we show the results of the excitation energies $\epsilon_k$ as a function of the wavevector $k$. Such energies are calculated from the generalization of Eq.~(\ref{GAP}) to values of $k$ away from the minimum
\begin{equation}
E_k\left(\frac{N}{2}+1, \frac{N}{2}\right)=E\left(\frac{N}{2},\frac{N}{2}\right)+\mu + \epsilon_k \;.
\label{EXC}
\end{equation}
The left hand side of the above equation is the energy of the polarized system with $N_t=N_b+1$, which contains a single unpaired particle placed in the plane-wave state with wavevector ${\bf k}$ complying with PBC. The definition (\ref{EXC}) of the excitation energy $\epsilon_k$ coincides with the one of the quasiparticle energy (\ref{Dispersion}) derived from BCS theory.

The result in the BCS regime ($k_F\lambda=0.5$) is shown in Fig.~\ref{fig9} and in the BEC regime ($k_F\lambda=0.25$) in Fig.~\ref{fig10}. In both cases the calculations are performed for $N_b=29$ and $N_t=30$ and $\epsilon_k=E_k(30,29)-E(29,29)-\mu$, where $\mu$ is the FN-DMC result of the chemical potential obtained from Eq.~(\ref{GAP}). From Fig.~\ref{fig9} one can see that the excitation spectrum has a minimum at $k\simeq k_F$. The blue double arrow shows the FN-DMC value of $\Delta_{gap}$ as obtained from Eq.~(\ref{GAP}). The dashed line is the expression (\ref{Dispersion}) for the dispersion relation, where for $\mu$ and $\Delta$ we use the FN-DMC results. Compared to the FN-DMC excitation energies, the minimum of (\ref{Dispersion}) is significantly shifted towards a larger value of $k$. We interpret this fact as the effect of the intralayer interactions which renormalize the mass of the quasiparticles. The solid line in Fig.~\ref{fig9} is the modified dispersion relation
\begin{equation}
\epsilon_k = \sqrt{\left( \frac{\hbar^2 k^2}{2m^{\star}}-\mu\right)^2 + \Delta^2} \;,
\label{Dispersion_1}
\end{equation}
where one accounts for the effective mass $m^{\star}$ that is treated as a fitting parameter. Fig.~\ref{fig9} shows that the above expression well reproduces the FN-DMC spectrum. The extracted value of the effective mass is $m^{\star}/m=0.77(3)$. Fig.~\ref{fig10} shows the results of the excitation energy in the BEC regime. In this case both the FN-DMC results and the BCS Eq.~(\ref{Dispersion}) (dotted line) exhibit a minimum at $k=0$. Furthermore, a good agreement is found for all wave vectors.

\section{Conclusions}

We investigated the superfluid state of a one-component gas of dipolar fermions in a bilayer configuration using the FN-DMC method. We calculated the ground-state energy, the superfluid gap, the chemical potential and the excitation spectrum as a function of the distance between the two layers. Comparison is made with the results of a simple mean-field theory valid in the low-density limit where, in particular, in-plane interactions are completely neglected. We find
that the equation of state and the superfluid gap exhibit a novel type of crossover from a BCS to a BEC regime as a function of the interlayer distance. In contrast to the more standard BCS-BEC crossover in two-component Fermi gases with resonantly enhanced contact interactions, the in-plane repulsion and the long-range nature of the interaction play here an important role, which for high enough density can lead to the competition between fermionic superfluidity and crystallization~\cite{Matveeva12}.

\section{Appendix A. Treatment of the potential interaction energy}
Since the dipole-dipole force is long range, the potential energy contributions arising from in-plane $V_{b(t)}$ and interlayer $V_{int}$ interactions require a careful treatment. The in-plane contribution from the bottom layer is given by 
\begin{equation}
V_b=\sum_{i<i^{\prime}}^{N_b}\frac{d^2}{|{\bf r}_i-{\bf r}_{i^{\prime}}|^3} + \frac{1}{2}\sum_{i,i^{\prime}}^{N_b}\sum_{{\bf R}\neq0} \frac{d^2}{|{\bf r}_i-{\bf r}_{i^{\prime}}-{\bf R}|^3} \;,
\label{V}
\end{equation}
where $i$ and $i^{\prime}$ label particles of the bottom layer in the simulation cell and the vectors ${\bf r}_{i^{\prime}}+{\bf R}$ correspond to the positions of all images of particle $i^{\prime}$ in the array of replicas of the simulation cell. 
The contribution from the top layer $V_t$ has the same form as Eq.~(\ref{V}), where the projections of the positions of top-layer particles onto the bottom layer are taken and $N_b$ is replaced by $N_t$.
Similarly, the contribution from interlayer dipolar interactions is given by
\begin{equation}
V_{int}=\sum_{i,j}^{N_b, N_t}\sum_{{\bf R}}\frac{d^2 (|{\bf r}_i-{\bf r}_j-{\bf R}|^2 -2\lambda^2)}{(|{\bf r}_i-{\bf r}_j-{\bf R}|^2 +\lambda^2)^{5/2}}.
\label{Vint}  
\end{equation}
We calculate the mean interaction energy using a procedure that takes advantage of the fast $1/r^3$ decay of the dipole-dipole potential:
\begin{eqnarray}
\langle V\rangle&=&(V_b)_{R_{c_1}} + (V_t)_{R_{c_1}} + (V_{int})_{R_{c_2}}
\nonumber\\
&+& E^b_{\text{tail}_1} + E^t_{\text{tail}_1} +E_{\text{tail}_2}.
\label{Vaver}
\end{eqnarray}
Here $(V_{b(t)})_{R_{c_1}}$ and $(V_{int})_{R_{c_2}}$ denote the sums (\ref{V}) and (\ref{Vint})  with the constraints $|{\bf r}_{i,j}-{\bf r}_{i^{\prime},j^{\prime}}-{\bf R}|\le R_{c_1}$ and $|{\bf r}_i-{\bf r}_j-{\bf R}|\le R_{c_2}$, respectively. 
The corresponding tail contributions $E^{b(t)}_{\text{tail}_1}=\pi d^2 N^2_{b,t}/(R_{c_1}L^2)$ and $E_{\text{tail}_2}=2\pi d^2 N_b N_t R^2_{c_2}/[L^2(\lambda^2 +R^2_{c_2})^{3/2}]$ are obtained by assuming a uniform distribution of particles for distances larger than the cut-off range.

\section{Appendix B. Extrapolation to the thermodynamic limit}
\begin{figure}
\begin{center}
\includegraphics[width=8.0cm]{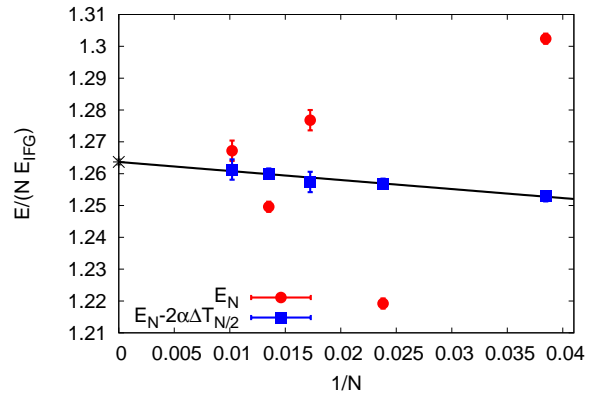}
\caption{(color online). Finite-size scaling for the case of the Jastrow-Slater wave function at $k_F\lambda = 0.6$. Red circles are the FN-DMC results for $N$ particles, blue squares  correspond to the values of energy corrected with the subtraction of $2\alpha \Delta T_{N/2}$, the black line is the linear fit of the form $E_{TL} + \frac{\beta}{N}$ and the black cross shows the extrapolated value $E_{TL}$.  
\label{fig11}}
\end{center}
\end{figure}

We are interested in the system properties in the thermodynamic  limit, therefore finite-size errors should be taken into account. In the case of the BCS-type wave function, we find that  the energy scales linearly as a function of $1/N$ and one can readily perform the extrapolation to $N\to\infty$. For the Jastrow-Slater wave function, instead, the extrapolation to the thermodynamic limit is made in a way similar to the case of the single-layer Fermi liquid~\cite{Matveeva12}. At each $k_F\lambda$ we perform simulations for $N/2 = 13, 21, 29, 37, 49$ all corresponding to closed-shell configurations for a two-dimensional Fermi gas. In order to obtain the energy in the thermodynamic limit $E_{TL}$ the following fitting formula is used
\begin{equation}
E_N=E_{TL}+ 2 \alpha \Delta T_{N/2}+ \frac{\beta}{N}, 
\label{E_TL}
\end{equation}
where $E_N$ is the FN-DMC energy for the system of $N$ particles and the fitting constants are $\alpha$ and $\beta$. Here $\Delta T_{N/2} = (N/2)E_{IFG} - T_{N/2}$ is the finite-size error in the energy of the noninteracting gas of $N/2$ particles, being $T_{N/2}$ the corresponding kinetic energy of $N/2$ particles. An example of finite-size dependence at $k_F\lambda = 0.6$ is shown in Fig.~\ref{fig11}. As one can see, the scattered distribution of energies for $N$ particles is largely suppressed once the corrections to the kinetic energy $2\alpha \Delta T_{N/2}$ are subtracted. The resulting energies (blue squares) linearly depend on $1/N$ allowing for a reliable extrapolation to the thermodynamic limit.

For a single-layer Fermi liquid the meaning of the coefficient $\alpha$ is the inverse effective mass of a quasiparticle~\cite{Matveeva12}. At $k_Fr_0=0.5$ its value is $m/m^\star=1.15(1)$. For the bilayer system we recover the same value of $\alpha$ at $k_F\lambda = 0.75$. This coefficient slightly changes for smaller distances: $\alpha = 1.21(1)$ at $k_F\lambda = 0.6$ and $\alpha=1.3(1)$ at $k_F\lambda = 0.5$ (notice that in the last case the Jastrow-Slater wave function already gives a higher energy than the BCS-type wave function).


\begin{thebibliography}{99}
\bibitem{Note} This value of the energy for dipolar bosons in a single layer was obtained performing DMC simulations as in Ref. \cite{Astra07}.
\end{thebibliography}
\end{document}